# Parallel Fourier Ptychography reconstruction

GUOCHENG ZHOU,[1] SHAOHUI ZHANG,[1,2,3] YAO HU,[1] LEI CAO,[1] YONG HUANG,[1,4] AND QUN HAO[1,2]

[1]*School of Optics and Photonics, Beijing Institute of Technology, Beijing 100081, China*
[2]*Yangtze Delta Region Academy of Beijing Institute of Technology, China*
[3]*zhangshaohui@bit.edu.cn*
[4]*huangyong2015@bit.edu.cn*

**Abstract:** Fourier ptychography has attracted a wide range of focus for its ability of large space-bandwidth-produce, and quantative phase measurement. It is a typical computational imaging technique which refers to optimizing both the imaging hardware and reconstruction algorithms simultaneously. The data redundancy and inverse problem algorithms are the sources of FPM's excellent performance. But at the same time, this large amount of data processing and complex algorithms also greatly reduce the imaging speed. In this article, we propose a parallel Fourier ptychography reconstruction framework consisting of three levels of parallel computing parts and implemented it with both central processing unit (CPU) and compute unified device architecture (CUDA) platform. In the conventional FPM reconstruction framework, the sample image is divided into multiple sub-regions for separately processing because the illumination angles for different subregions are varied for the same LED and different subregions contain different defocus distances due to the non-planar distribution or non-ideal posture of biological sample. We first build a parallel computing sub-framework in spatial domain based on the above-mentioned characteristics. And then, by utilizing the sequential characteristics of different spectrum regions to update, a parallel computing sub-framework in the spectrum domain is carried out in our scheme. The feasibility of the proposed parallel FPM reconstruction framework is verified with different experimental results acquired with the system we built.

## 1. Introduction

Fourier ptychography microscopy (FPM) [1-4] has attracted a wide range of interest for its high space-bandwidth product and quality phase imaging ability. It is a typical computational imaging approach, which refers to joint optimization of both hardware and algorithms to increase the imaging performance or achieve functions that cannot be realized by conventional microscopic platforms. In conventional microscopy, limit by numerical aperture (NA) of the objective, the spatial resolution is always insufficient because only small part of the diffraction light emitted from sample can enter the imaging optics and participate in imaging. In FPM platforms, a programmable LED array is usually utilized as light source to provide angular varied illuminations for the sample under test to shift its spatial spectrum. A set of low-resolution (LR) intensity images corresponding to different illumination angles are stitched together in the Fourier domain through phase retrieval [5] and aperture synthesis [6]. FPM is a kind of super-resolution approach, in which the spatial resolution is evidently improved while the field of view (FOV) unchanged. Nevertheless, the performance improvement comes at the cost of a very large amount of data redundancy and calculation complexity. The FPM algorithms framework is an iterative optimization procedure, which refers to alternating projections [5, 7] onto spatial and frequency domain constraints.

Fast Fourier transform (FFT) and inverse fast Fourier transform (IFFT) are utilized repeatedly in FPM to impose the spatial module constraints and frequency support constraints. The spatial domain module constraints indicate that the amplitude part of the sample complex distribution in the spatial domain should be consistent with the LR intensity images recorded by the camera, while the frequency domain support constraint means the circular low-pass filter determined by the NA of the objective and the wavelength of the illumination light. The above-mentioned iterative procedure is computationally intensive and time-consuming, which limit the speed of data processing and result display. Different kinds of approaches have been taken to deal with the problem of large data volume and calculation volume in the FPM algorithm framework, such as using better optimization strategies [8, 9], optimizing illumination patterns to reduce data redundancy [10-12], and utilizing neural network to replace the conventional algorithm framework [13, 14] et.al. Nevertheless, the above-mentioned approaches are not trouble free. For example, changing the optimization strategy has a relatively limited increase in data processing speed, while reducing data redundancy has a certain impact on the recovery results compared with the conventional FPM, and the principle of data processing methods using neural networks is not clear, and generalization performance is difficult to guarantee. Therefore, an acceleration approach without changing the inverse problem itself is very necessary for FPM, which can be also applied to more fields. What's more, one feature of the computational imaging is to make fully use of the computing power to mine the hidden associations between data, and reducing the hardware requirements of the optical imaging systems. Therefore, a more general acceleration strategy is in urgent to deal with increasing larger amounts of data and increase accuracy.

Parallel computing, a type of computation in which many calculations or processes are carried out simultaneously, has been applied to accelerate optical measurement and imaging techniques in the past few years [15-17]. Parallelism is the prerequisite for a certain problem to be accelerated in parallel computing. A high parallelism can always be exploited in computational imaging data processing flow as long as the data have a certain level of independence in a certain domain. As a computational imaging approach, FPM has very high parallelism at the data level and the data collection process level. Therefore, constructing a reasonable parallel computing framework will be of great help to increase the speed of FPM data processing, and to expand its potential applications range. In fact, although the concept of parallel computing is intuitive and simple, the implementation may not be easy because of the hardware platform construction and algorithm code programming. Thanks to the emergence of general software and hardware framework, such as OpenCL and compute unified device architecture (CUDA), it becomes easier to implement parallel acceleration for specific optical measurement or imaging problems.

In this paper, to realize parallel computing framework for FPM, we will fully explore the parallel features in the data processing process and implement the parallel acceleration framework based on multi central processing units (CPUs) and CUDA. We demonstrate the feasibility of the proposed method with single CPU, multi CPUs, single graphics processing unit (GPU) respectively. This paper is organized as follows. The system structure and algorithm principle of FPM is presented in Section 2. The parallelism of FPM algorithm and parallel computing framework are presented in detail in Section 3. The processing results of different strategies will be shown in Section 4. Finally, discussions and conclusions are summarized in Section 5.

## 2. Principle

### 2.1 FPM principle and algorithm

The working flowchart of FPM is presented in Fig. 1, in which the sample is illuminated sequentially by a set of angular varied monochromatic plane wave instead of a conventional specific illumination mode, such as Kohler light source. The raw data consists of the LR intensity images corresponding to each illumination angle. In general, the limited objective lens manufacturing process level results in a comprise between achievable imaging field of view (FOV) and spatial resolution. Therefore, an objective with relatively low NA is always utilized to obtain a large FOV and achieve high spatial resolution by FPM framework.

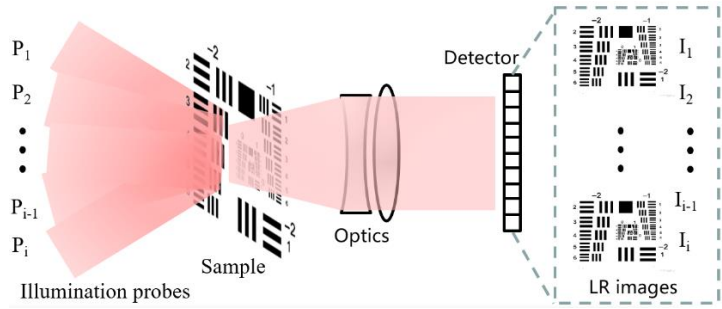

Fig.1 workflow of the FPM

FPM shares roots with phase retrieval in coherent diffraction imaging (CDI) [18, 19] and synthetic aperture imaging concept in astronomy. As shown in Fig.2, by iteratively recover the phase part and stitches the captured LR intensity images together in the Fourier domain, a large FOV and high spatial resolution complex image can be obtained. The core of procedure in this algorithm is the alternating projection [7], a simple optimization approach for computing a point in the intersection of some convex sets using a sequence of projections and constraints onto the sets. Nevertheless, instead of two fixed sets, the sets in FPM framework corresponds to low pass circular filters in the Fourier domain and sets of LR intensity images in the spatial domain, and that is the reason for its rapid convergence. In each iteration optimization loop, steps shown in the middle square in Fig.2, information of LR images corresponding to every illumination angle are utilized to constraint the HR complex image distribution. And, it usually needs several iterative optimization loops to get a relatively good reconstruction HR complex image.

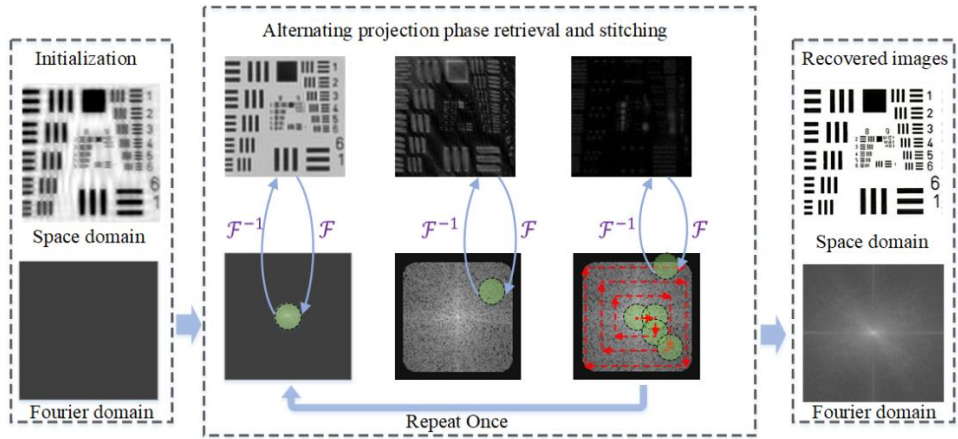

Fig.2 FPM algorithm flowchart

## 2.2 Typical FPM system implementation

In order to reduce the system complexity or construction cost without significantly affecting the imaging quality, a lot of approximate methods are used in the actual construction of the FPM systems. In the actual system construction and construction process, a programmable LED array is always used to provide the angle varied illumination. Nevertheless, LED unit is more likely to be a point light source, and its luminous wave front is closer to a spherical wave rather than a plane wave. As shown in Fig.3, a programmable LED array is used as light source for its low cost and flexibility of manipulation. Fig.3 (a) indicates that LED units located at different positions can provide different illumination angles, which are determined by the sample's height and the positions in the two-dimensional coordinate system constructed by the

LED board. Fig.3 (b) shows a commercially available LED array, which can realize flexible spatial and sequential lighting modes through Raspberry Pi or Microcontroller Unit (MCU), et.al. A critical prerequisite for the LED array to be used as the FPM illumination source is that the illumination light onto the sample can be approximated as a plane wave, which demands the size of the sample of interest should be much smaller than the distance between the LED unit and the sample. Therefore, in the actual FPM reconstruction process, the sample is usually divided into a number of small areas that meet the above approximation and processed separately, and then stitched together to obtain a complete HR image.

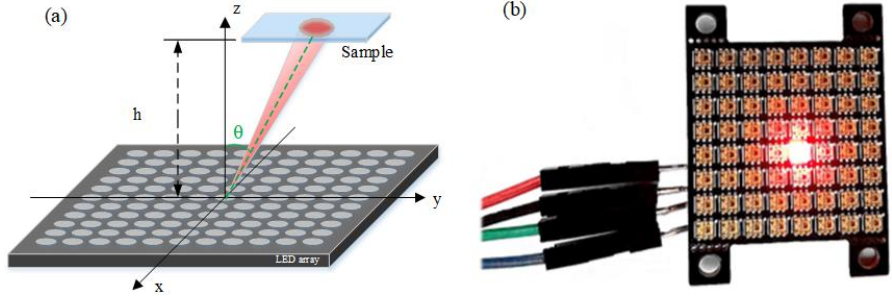

Fig.3 Programmable LED array for FPM illumination. (a) Schematic of single LED illuminating the sample, (b) LED array board photo

In addition to the requirement of approximating spherical wave emitted from LED to a plane wave, the change of the illumination angle for sub-areas located at different positions is also the reason why we must partition the samples and reconstructed them separately. Fig.4 indicates the difference with plane wave illumination and LED illumination. The illumination angle for each sub-area of the whole sample is the same for parallel plane wave illumination as shown in Fig.4 (a), while the incident illumination angle in different sub-areas vary with the spatial positions for LED illumination as shown in Fig.4 (b). Therefore, we have to divide the sample into several sub-areas with different illumination angles, and processed them separately. Furthermore, due to the non-planar distribution and the non-ideal position/posture characteristics as mentioned in [20], different subregions will include different defocus distances, which also indicates that dividing the sample into several sub-areas is important to extending the depth of field (DOF) in FPM.

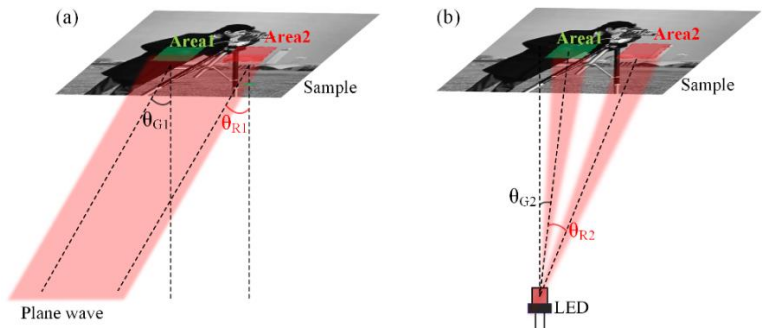

Fig.4 Illumination angles corresponding to different illumination modes. (a) with parallel plane wave, and (b) with LED.

## 3. Parallel FPM reconstruction implementation

In order to accelerate the FPM data processing speed and therefore expand its potential application range, we will fully investigate the parallelism in the FPM algorithm framework and build a general parallel computing framework to realize fast HR complex image reconstruction. The parallelism of each algorithm depends on its data structure and data processing flow. Similar to ptychography iterative engine (PIE) [21], FPM algorithm framework relies on complex-valued matrix operations and the Fourier transforms. The parallel computing strategy consists of three types of parallel operations, and it will be written in Python by using single CPU, multi-CPUs and single GPU as comparisons.

*3.1 Spatial parallel computing*

The spatial parallel computing framework is designed and constructed based on the data independence characteristics of the sample, as discussed in session 2.2. As shown in Fig.5 (a), the sample is divided into multiple sub-areas that partially overlap each other. For each small region, the spherical wave emitted by the LED can be approximated as a plane wave illumination. The overlap between each adjacent sub-region is used to eliminate the discontinuities in both amplitude and phase domain that may occur in the process of stitching sub-regions back into a complete large FOV image [17, 22-24]. Taking four sub-regions as an example, Fig.5 (b) indicates how the data is processed separately and simultaneously. Each sub-region is an image with pixel size 256x256, and allocated to one CPU for iterative optimization operations as described in section 2.1, without the need for data sharing or synchronization between each other. The overlap range between adjacent images is set to 256x26 pixel size in our experiments.

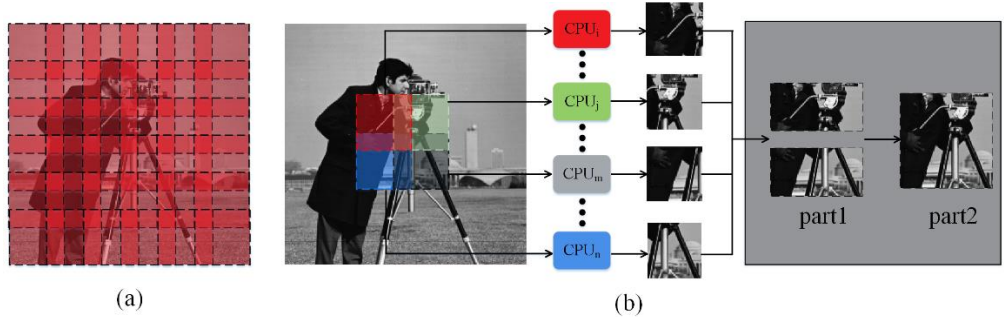

(a)  (b)

Fig.5 Schematic diagram of spatial parallel computing FPM algorithm

When the phase recovery and HR reconstruction for each sub-region shown in Fig.5(b) are accomplished on each CPU in parallel, they can be stitched back to a whole complex image with large FOV and high spatial resolution. The discontinuities appeared both in spatial and phase parts between adjacent sub-regions can be eliminated effectively by appropriate image stitching methods.

$$f(x,y) = \begin{cases} f_1(x,y), & \text{if } x < (row - \dfrac{overlap}{2}) \text{ for part 1} \\ & \text{or } y < (col - \dfrac{overlap}{2}) \text{ for part 2} \\ \dfrac{\mu_1}{\mu_2} \cdot f_2(x,y), & \text{otherwise.} \end{cases} \quad (1)$$

As shown in Eq. (1), the complex HR images are stitched by employed a method similar to [17], where $f(x,y)$ indicates the final stitched HR complex image. $f_1$ and $f_2$ correspond to two HR sub-images prepared for image stitching, and $(x,y)$ is the two-dimensional (2D) coordinates. The weight coefficients $\mu_1$ and $\mu_2$ are the complex-valued means corresponding to $f_1$ and $f_2$. "*row*" and "*col*" indicate the width and height of image size respectively, and "*overlap*" is set to 26, which is 10% of the image pixel size in our experiment. For instance, the pixel size of input images $f_1$ and $f_2$ is 256x256, thus, the output image size of $f(x,y)$ will be 486x256 for part1 as shown in Fig. 5. And in turn, this output will be as the input for part2 for image stitching in Fig. 5.

### 3.2 Pipelined parallel computing

Besides the data parallelism in FPM datasets, the data processing procedure that each LR intensity image is utilized sequentially also provides the possibility of parallel acceleration. Fig. 6 show two typical FPM HR spectrum recovery routines and the corresponding parallel computing implementation methods. Fig.6 (a) presents the most common spiral outward update routine, while Fig.6 (c) shows a raster scanning update routine. Fig.6 (b) and (d) are the corresponding pipelined parallel computational approaches, whose main idea is processing different sub-spectrum regions in different iteration loops in parallel. The core skill of the above pipelined parallel computing implementation frameworks is how to adjust the update sequence reasonably so that the next update iteration loop will not affect the previous one.

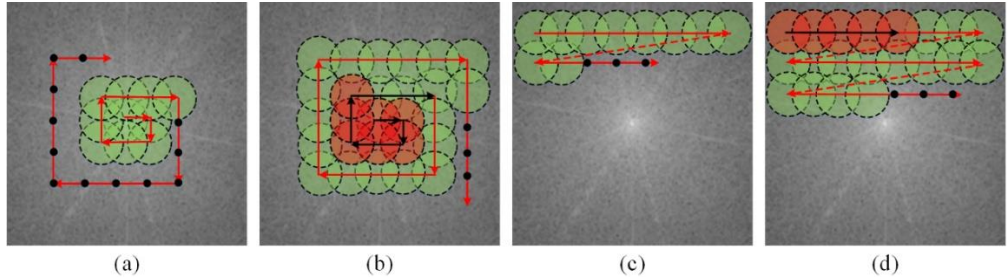

Fig.6 Different FPM recovery routines and corresponding parallelism implementations

### 3.3 Parallel FFT and IFFT

Besides the data and data processing procedure, the specific data processing operation can also have some degree of parallelism. As mentioned in section 2.1, the FPM reconstruction algorithm consists of iterative 2D FFT/IFFT and matrix multiplex operations. The calculation amount of FFT/IFFT depends heavily on the data size, referring to the total pixel number of each divided sub-image in section 3.1.1. Although the sample has already been divided and processed separately according to the spatial parallel computing strategy, the calculation amount for the reconstruction of each sub-image can still be accelerated by investigating the parallelism of 2D FFT/IFFT. Fortunately, there exists a well-developed parallel version FFT/IFFT in CUDA platform named Pytorch in Python, and cuFFT/cuIFFT in C++, which can evidently save the time of 2D Fourier transform and inverse Fourier transform.

### 3.4 The hybrid parallel computing framework

The flow chart of proposed method consists of three parts, the LR images segmentation, the HR sub-regions reconstruction and the HR complex images stitching. As shown in Fig.7, depending on whether the data acquisition process is included, the whole parallel processing framework can be classified as online testing and offline testing. In the former one, LR images will be processed one by one, indicating that once a LR image is captured by camera, it will be segmented into several sub-regions, reconstructed by FPM method corresponding to the specific illumination angle, immediately. Since changing the LED illumination pattern and capturing LR image also need a certain amount of time to complete (for example, the typical value in our experiment is 0.3s depended on our devices), parallelizing image acquisition and data processing can significantly improve the overall efficiency of the FPM system. Thus, the whole processing time can be reduced effectively due to the mining of more parallelism. How many iterations will be conducted during the acquisition processing loop depends on the time-consuming proportional relationship between hardware acquisition and one iteration of the optimization algorithm. At last, a whole HR images can be obtained at the same time as the end of the last LED illumination mode. Nevertheless, considering the non-planar distributions and non-ideal position/posture of the sample as mentioned above, all or part of FOV will be defocused when LR images acquisition. Thus, digital refocusing is important in FPM to extend the DOF. According to [20], digital refocusing relies on at least 9 LR images corresponding to BF images as prerequisite, and needs utilized before HR reconstruction, the on-line parallel framework is not suitable for this situation. The offline parallel framework which process all the LR images simultaneously can accomplish such kind of tasks. As shown in Fig.7(b), the offline strategy refers to the defocus distance calculation and HR reconstruction process will be implemented after all LR images have been acquired. In fact, if the image acquisition speed of the hardware system is very fast, then in order to reduce the complexity of the programming, we can only use this offline parallel HR reconstruction framework. The details of processing flow corresponding to online and offline testing are shown in Fig. 7.

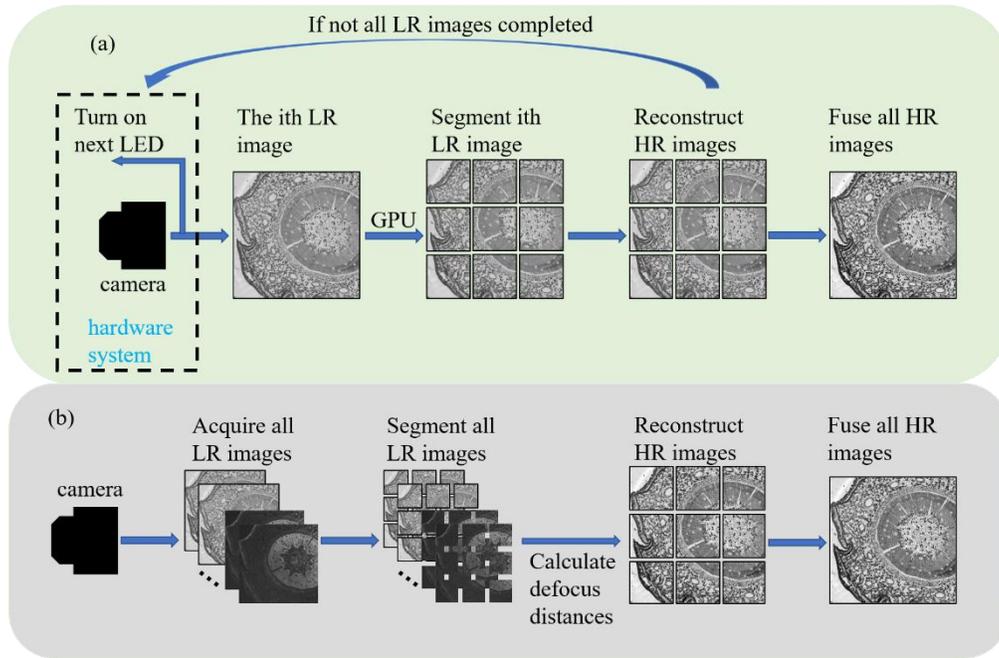

Fig. 7 Processing flows corresponding to different proposed strategies. (a) On-line testing strategy. (b) Offline testing strategy.

As shown in Fig. 7 (a), when a LR image is acquired by CPU, it will be sent to GPU memory immediately. And then, segmenting and reconstructing processing will be implemented sequentially. At the same time, due to the CPU working is completely parallel with GPU working, thus, the LED pattern will be changed to next pattern by CPU. Generally, LED pattern changing needs 0.3s by our device, which is longer than the cost time of GPU working (typical value is 0.1s for 16 subregions corresponding to each LR image). Thus, the whole time of on-line testing is almost depended on the time of LED pattern changing. Finally, if all LR images are acquired and processed, all complex HR sub-images will be stitched together to obtaining a FOV HR images.

As shown in Fig. 7 (b), in offline strategy, all LR images will be acquired firstly, and they can be sent to GPU immediately or only processed in CPU. And then, segmenting, defocus distance calculating, and reconstructing processing are implemented sequentially for all LR images. Taking CPU processing as an example, as mentioned in Fig. 5, each sub-region will be allocated a CPU for calculation. In Fig. 7 (b), there are 9 sub-regions as an example, thus, totally 9 CPUs will be allocated for calculation processing. And finally, after all LR images are reconstructed in HR images, image stitching strategy is implemented to obtain a FOV complex HR image. Different from the online testing, offline testing acquires all LR images firstly, which means acquiring process cannot parallel with reconstructing process. Although it will increase the whole processing time, it is suitable for releasing as a procedure for other researches.

## 4. Experiments and results

To demonstrate the feasibility of the proposed method, we build an FPM system as shown in Fig. 8. The system setup of FPM consists of a camera (FLIR, BFS-U3-200S6M-C, sensor size 1'', dynamic range 71.89dB, pixel size 2.4μm) for LR image acquisition, a LED array (CMN, P 2.5, 64×64) for varied angle illumination, and an objective (2X, NA≈0.1). The distance between sample plane and LED array is set to 83mm. The exposure time of camera is set to 30ms. It's worth noting that, to increase the acquisition time as much as possible, the acquisition mode of camera is set to external triggering. Then, as mentioned above, we acquire a set of LR images, segment them into several sub-regions and reconstruct them respectively. The pixel size of each subregion is set to 256x256, which is a suitable size for digital refocusing and FPM reconstruction. The overlap between adjacent subregions is set to 256x26, which is approximately 10% of the size of subregion. And finally, several complex HR images corresponding to different subregions will be stitched together using the image stitching method mentioned in Eq. (1) to obtaining a FOV HR image.

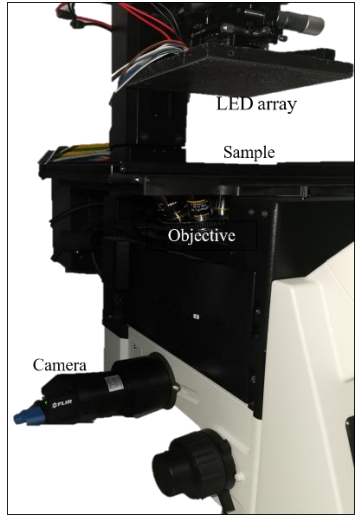

Fig. 8 System setup of conventional FPM.

### 4.1 Processing speed

Before comparing the different processing speed with single CPU, multi-CPUs and single GPU in our method, we use a USAF test chart to demonstrate the feasibility of parallel reconstruction of our method with multi-CPU firstly. The CPU used in our experiment is $8^{th}$ Gen Intel(R) Core (TM) i5-8300H, which has 8 threads totally. According to Fig. 7 (b), we acquire 169 LR images, and then, segment each LR image into 4 sub-regions whose pixel size is set to 256x256, and the overlap region is set to 256x26. After calculating defocus distances corresponding to different sub-regions and HR reconstruction process, the reconstructed HR amplitude of each sub-regions are shown in Fig. 9 (a1-a4). Fig. 9 (b1) is the enlarge subregion in Fig. 9 (a4). Besides, to verify the feasibility of digital refocusing in our method, a result corresponding to conventional FPM strategy is shown in Fig. 9 (b2) as a comparison. Fig. 9 (c1) is the LR BF image corresponding to Fig. 9 (a4), and Fig. 9 (c2) is the enlarged sub-region in Fig. 9 (c1). Generally, the results shown in Fig. 9 can verify the parallel HR reconstruction feasibility of our method.

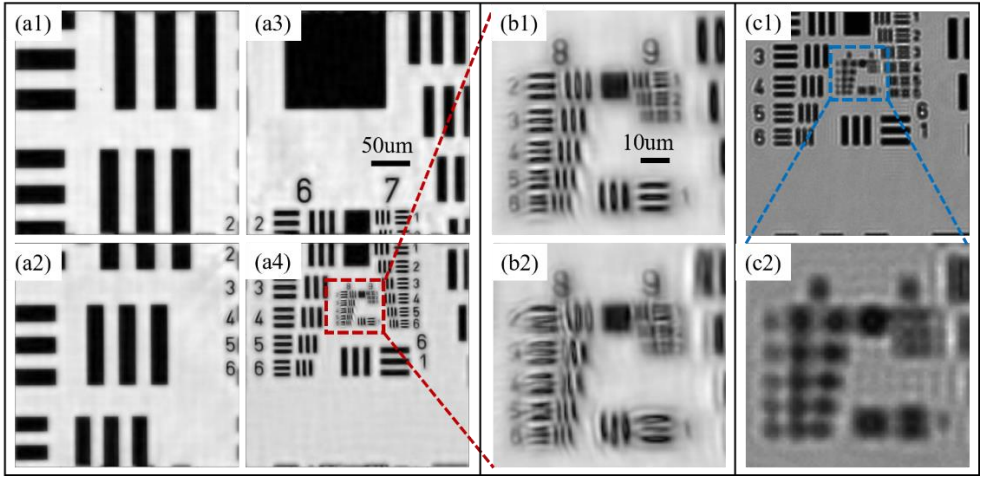

Fig. 9 HR reconstructed amplitude of USAF corresponding to different sub-regions. (a1-a4) HR reconstructed amplitude of different sub-regions. (b1) enlarged sub-region corresponds to (a4). (b2) HR reconstructed amplitude without digital refocusing. (c1) LR BF image corresponds to (a4). (c2) enlarged sub-region corresponds to (c1).

To compare the processing speed of single CPU, multi-CPUs, and single GPU, the experiments are carried out according to the strategy shown in Fig. 7 (b). The GPU (Nvidia GeForce GTX 1050 laptop GPU) is used in this experiment. As shown in Fig. 7 (b), we acquire 169 LR images firstly. And then, each LR image is divided into 16 sub-regions, in which the pixel size is set to 256x256. After calculating defocus distances corresponding to different sub-regions, HR reconstruction will be implemented. The total iteration number is set to 1 in our experiment. We compare the different processing times of different number of sub-regions corresponding to different strategies, respectively. Furthermore, as mentioned above, each LED pattern changing costs 0.3s. If the exposure time of the camera is set to 0.03s, the total acquisition time is approximately 56s. We are noting that, according to Fig. 7 (a), due to each LR image (16 sub-regions as an example) needs only 0.1s for processing, the reconstruction process can be parallel to acquisition process in the online strategy. Thus, the costing time of online strategy is 56s as long as the LR image acquisition is finished. However, by using the offline strategy, the acquisition process and reconstruction process are implemented sequentially. The reconstruction processing times corresponding to single CPU, multi-CPUs, and single GPU are compared as follows.

As shown in Fig. 10, the processing time of one sub-region costs 6.7s for single CPU. Thus, with the increasing of sub-region number, the processing time increases almost linearly. By implementing the multi-CPUs processing, the processing times are 6.6, 7.8, 8.6 et. al., which increases slowly. Theoretically, the processing time will be same as long as the subregion number does not exceed the number of threads of CPU. We think each thread requires some time for starting, thus, causing this different phenomenon between theory and experiment. Furthermore, the GPU shows more better performance even only single GPU is used in our experiment as shown in Fig. 10. One subregion takes 1s for processing, and then, it is 1.9s, 2.8s et. al with the number of subregions increasing.

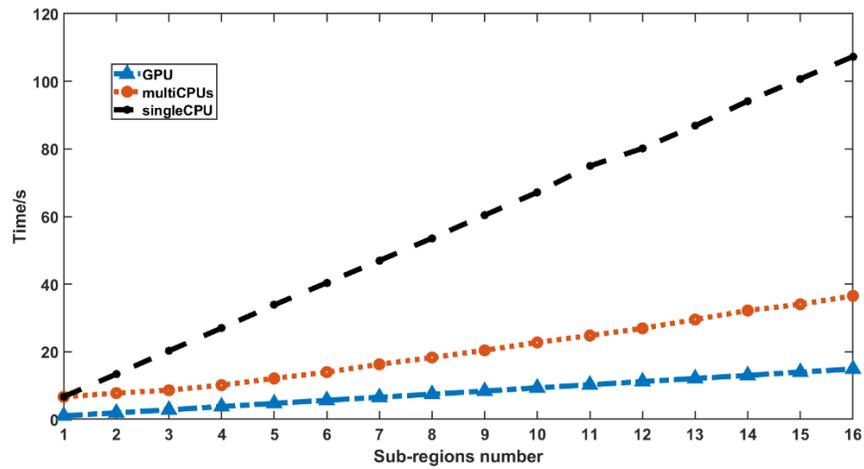

Fig. 10 Different processing time of single CPU, multi-CPUs and single GPU corresponding to different subregion numbers.

### 4.2 Image stitching results

We use a Bee wing biomedical sample as an example to verify the effectiveness of image stitching. Similar to Section 4.1, 169 LR images are acquired and each LR image is divided into 16 sub-regions. As shown in Fig. 11 (b1-b2), if the overlap regions are 0, the discontinuities between adjacent sub-regions cannot be eliminated, especially in the phase part. The value range of phase part corresponding to different sub-regions cannot be stitched in a suitable range, resulting in the discontinuities between adjacent sub-regions as shown in Fig. 11 (b2). Nevertheless, if the LR images have overlap area, the discontinuities between adjacent sub-regions can be eliminated well as shown in Fig. 11 (a2).

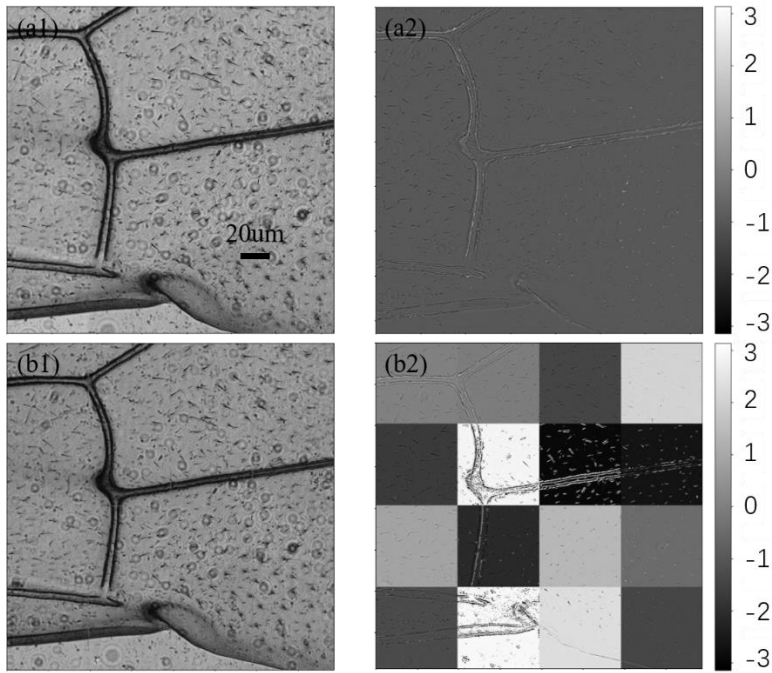

Fig. 11 Complex HR reconstructed results of Bee wing. (a1) and (a2) are reconstructed amplitude and phase with overlap region between adjacent sub-regions. (b1) and (b2) are reconstructed amplitude and phase without overlap region between adjacent sub-regions.

## 5. Conclusion and future work

In this paper, we propose a parallel Fourier ptychography reconstruction framework with multi-CPUs and GPU. According to the reconstructed characteristics of FPM, we proposed three parallel strategies in our paper and integrated them into our parallel FPM reconstruction framework, including thread parallelism, pipelined parallelism and FFT/iFFT operation parallelism. Furthermore, in order to eliminate the discontinuities between adjacent sub-regions causing by image stitching, overlapping region is retained when image segmenting. And then, through some image stitching methods, the discontinuities can be eliminated. The experimental results indicate that both multi-CPUs strategy and single GPU strategy can increase the image processing speed compared to single CPU strategy. And the performance of single GPU strategy is also better than multi-CPUs strategy even there is only one GPU in data processing, which shows the remarkable performance of GPU in image processing, especially in FFT and iFFT.

Although the results have demonstrated the feasibility of the proposed method, there are still some work can be improved in the future. For instance, limited by the working characteristic of GPU, it can only process one task each time, which means different subregions are actually processed one by one, the processing speed is almost improved based on improving the speed of FFT and iFFT. In the future work, multi-GPUs strategy can be implemented in FPM reconstruction. The data can be prepared in CPU firstly, and then, different subregions' data will be sent to different GPUs for data processing, thus, realizing parallel data processing. Secondly, as mentioned in Section 4.1, the whole processing time of on-line testing strategy is depended on both processing time (using GPU or CPU) and LED changing time. Thus, the lesser LED pattern changing, the faster processing speed. Actually, some researches have devoted to multiplexed coded illumination in FPM [11], which can be used in our work to reduce the acquisition time.

**Disclosures.** The authors declare no conflicts of interest.

**Data availability.** Data underlying the results presented in this paper are not publicly available at this time but may be obtained from the authors upon reasonable request.